\RequirePackage{fix-cm}
\documentclass[11pt]{article}
\usepackage{graphicx}
\usepackage{multirow}
\usepackage[T1]{fontenc}
\usepackage{booktabs}
\usepackage{natbib}
\begin{document}

\title{Evidence for electro-induced membrane defects assessed by lateral mobility measurement of a GPi anchored protein
}


\author{Jean Michel Escoffre$^{1}$, Marie Hubert $^{1}$, Justin Teissi\'{e} $^{1}$, \\Marie Pierre Rols $^{1}$ and Cyril Favard $^{2}$ \\ $^{1}$ Institut de Pharmacologie et de Biologie Structurale,\\ CNRS UMR 5089,
31077 Toulouse Cedex, France\\$^{2}$Institut Fresnel, CNRS UMR 6133,
13397 Marseille Cedex, France}



\date{Submitted to Eur. Biophys. J. 29/11/2013}

\maketitle

\begin{abstract}
Electrotransfer is a method by which molecules can be introduced in
living cells through plasma membrane electropermeabilization. Here,
we show that electropermeabilization affects the lateral mobility of
RAE-1, a GPi anchored protein. Our results suggest that 10 to 20
$\%$ of the membrane surface is occupied by defaults or pores and
that these defaults propagates rapidly (<1 min) over the cell
surface. Plasmid DNA (pDNA) electrotransfer also affects the lateral
mobility of RAE-1. Furthermore, we clearly show that once inserted
into the plasma membrane, pDNA is totaly immobile and excludes
RAE-1, indicating that pDNA molecules are tightly packed together to
form aggregates at least in the outer leaflet of the plasma
membrane.

\end{abstract}

\section{Introduction}
\label{intro} The permeability of a cell membrane can be transiently
increased by applying external electric field pulses. This
phenomenon, called electropermeabilization or electroporation, leads
to the formation of "membrane defects" or "electropores" in the cell
membrane \citep{Neu89,Cha92,Wea93}. Since it is an elegant way to
introduce exogenous molecules into the cytoplasm, this method is
routinely used in basic research, as a very efficient way for the
delivery of drugs, oligonucleotides and plasmids (pDNA) but also
in-vivo for clinical applications
\citep{Mir06,Dau08,Low09,Esc12,Wei13}.
 Electropermeabilization is a
multi-step process occuring on different time scales \citep{Tei05}:
\begin{enumerate}
\item \textit{Induction step \textbf{(ns)}.} The electric field induces the membrane potential difference to increase.
When it reaches a critical value (about 200 mV) local transient
permeant structures appears.
\item \textit{Expansion step \textbf{($\mu$s)}.} Defects expend as long as the field, above the critical value, is
present.
\item \textit{Stabilisation step \textbf{(ms)}.} As soon as the field intensity is lower than
the critical value, a stabilisation process takes place within a few
milliseconds, which brings the membrane to the "permeabilized
state".
\item \textit{Resealing step \textbf{(s, min)}.} A slow resealing of the defaults is then occurring.
\item \textit{Memory effect \textbf{(h)}.} Some changes in the membrane properties remained present
on a longer time scale but the cell behavior is finally back to
normal.
\end{enumerate} If the kinetics of
electropermeabilization seems to be well established, very few is
known about the changes occurring at the cell and membrane molecular
levels \citep{Tei05}. Nevertheless, although structural changes in
the plasma membrane (i.e., formation of "membrane defects" or
"electropores") have never been directly visualized under the
microscope, other techniques have been used to observe
electropermeabilization. These include measurements of conductivity
of cell suspensions and pellets \citep{Kin79,Abi94,Pav05,Pav07},
electro-optical relaxation experiments on lipid vesicles
\citep{Kak96,Gri02}, charge pulse studies on lipid bilayers
\citep{Gri02,Ben79}, measurements of membrane voltage on cells with
potentiometric fluorescence dyes \citep{Hib93}, and monitoring the
influx or efflux of molecules and fluorescent dyes
\citep{Gab97,Gab99,Puc08,Rol98,Mir88,Tek94,Pra94,Pra95}. More
recently, molecular simulation on pure lipid models showed the
possibility of pore formation during the pulse and pore evolution up
to tenth of ns time scale \citep{Lev10,Tar05,Tie04}. Finally,
numerical computation of the evolution of theses pores has
authorized a more sophisticated, but still theoretical, description
of the phenomenon \citep{Kra07}. On pure lipid models, Krassowka et
al. predict mean size of "small" pores to be around 1 nm for 97\% of
them while mean size of "large" pores was around 20 nm with some of
them as big as 400 nm on 50 $\mu$m vesicles pulsed with a 0.6
$kV.cm^{-1}$ electric field intensity for 1 ms. While small
molecules (i.e., < 4 kDa) cross the permeabilized cell membrane
directly mainly by post-pulse diffusion, plasmid DNA (pDNA) first
interact with the electropermeabilized part of the membrane as shown
by the formation of localized aggregates \citep{Esc11,Puc08}. Taken
into account the size of the pDNA (i.e., 3 MDa, 30 nm in diameter)
and the negative charges of pDNA (as dielectric exclusion must also
be overcome) and supposing that the permeabilization is due to
conducting defects called "pores", then these membrane structures
must be large and stable \citep{par69}. Nevertheless, since the cell
membrane has a much more complex organization than a model lipid
bilayer, one expects that the location of regions where pDNA
electrotransfer occurs, will be determined not only by the local
electric field but also by the local membrane composition and
tension \citep{Ros11}. In order to sense the effect of
electropermeabilization on the plasma membrane, in the absence or
presence of pDNA, we have monitored the lateral mobility of a GPi
anchored protein Rae-1 during the resealing step, by means of
fluorescence recovery after photobleaching (FRAP) experiments. As a
GPi anchored, Rae-1 (MHC-I homologous protein) is located in the
outer leaflet of the plasma membrane without any partioning into
lipid domains previously described \citep{Nomura1996,Zou1996}.
Therefore GPi anchored has been considered as a good candidate to
report changes in the plasma membrane lateral state and in its close
interface with the outside of the cell. By measuring its mobility
before and after application of permeabilizing electric field pulses
in absence of pDNA, we reported a drastic and significant increase
of the half-time of fluorescence recovery and a decrease of the
mobile fraction, both at the anode and cathode facing pole of the
cell. These experimental data are in favor of the creation of
obstacles in the plasma membrane (e.g., "pores", "membrane
defects"). We then showed that when pDNA is inserted into the
membrane after electropermeabilization, Rae-1 is totally unable to
re-enter the area occupied by pDNA, confirming the direct
observation that pDNA is accumulated in a tightly bound manner into
clusters.

\section{Material and Methods}
\label{sec:1}
\subsection{Expression of Rae-1 in CHO cell line}
\label{sec:2} The eGFP-Rae-1 CHO cells have been generously made by
Dr. B. Couderc (EA3035, Institut Claudius Regaud, France). CHO cells
have been transfected by pDNA encoding Rae1-eGFP fusion protein
(generous gift from Dr. A. Aucher and Dr. D. Hudrisier, IPBS-CNRS,
UMR5089, France). The transfected cells are cultured under selective
pressure with G418 (1 $\mu$g/$\mu$L) (InvivoGen, San Diego, CA). The
eGFP-Rae-1 expressing cells were sorted out by flow cytometry
(FAScan; Beckman. Instruments, Inc. Fullerton, CA). Cells were then
grown as previously described \citep{Phe05}.
\subsection{pDNA labeling for electropermeabilization}
4.7-kbp plasmid (pEGFP-C1, Clonetech, Palo Alto, CA) carrying the
green fluorescent protein gene controlled by the CMV promoter was
prepared from Escherichia coli transfected bacteria by using
Maxiprep DNA purification system (Qiagen, Chatsworth, CA, USA). They
were covalently labeled with Cy-3 fluorophore using Label-IT nucleic
acid labeling kit (Mirus, Madison, WI, USA) according to the
manufacturer protocol. The fluorescent labeling did not affect the
function of expression cassette.
\subsection{Electropermeabilization}
Electropulsation was carried out with a CNRS cell electropulsator
(Jouan, St Herblain, France), which delivers square-wave electric
pulses. An oscilloscope (Enertec, St. Etienne, France) was used to
monitor the pulse shape. The electropulsation chamber was built
using two stainless-steel parallel rods (diameter 0.5 mm, length 10
mm, inter-electrode distance 5 mm) placed on a Lab-tek chamber
\citep{Maz09}. The electrodes were connected to the voltage
generator. A uniform electric field was generated. The chamber was
placed on the stage of the confocal microscope (Zeiss, LSM 510,
Germany). Electropermeabilization of cells was performed by
application of millisecond electric pulses, conditions required to
efficiently transfer macromolecules such as pDNA into cells
\citep{Rol98}. Ten pulses of 5 ms duration and 0.6 kV/cm amplitude
were applied at a frequency of 1 Hz at room temperature. For FRAP
experiments, the eGFP-Rae-1 CHO cells were seeded on a microscope
glass coverslip chamber
 (Labtek II system, NuncTM, Denmark) at 0.5.10$^6$ cells per well 24h before electropulsation.
 Cells were electropulsed in 200 $\mu$L of pulsation buffer (10 mM K$_2$HPO$_4$/KH$_2$PO$_4$, 1 mM MgCl$_2$,
 250 mM sucrose, pH 7.4). In absence of pDNA, the cell electropermeabilization was monitored
 by adding propidium iodide at 100 $\mu$M in the pulsation buffer. Eighty percent of cells located between
  the electrodes were permeabilized. The cell electropermeabilization in presence of Cy3-pDNA was
  performed after adding of 2 $\mu$g of Cy3-pDNA in 200 $\mu$l of pulsation buffer.
  Cy3-pDNA molecules interacted with more than 60\% of cells located between the electrodes.

\subsection{Fluorescence recovery after photobleaching}
FRAP experiments were conducted using a Zeiss LSM-510 confocal
microscope. The image sequence was acquired at a 5 Hz frequency
using the 488 nm line of an argon ion laser at a very low power to
avoid photobleaching during recording. After 50 images, 4 regions of
interests (ROI), of 1 $\mu$m radius each, which correspond to 0.86
$\mu$m waist of a Gaussian beam, located in front of anode and
cathode respectively and on left and right sides of the cell (see
~\ref{tab1} for illustration), were rapidly photobleached (t < 300
ms) at maximal laser power. Fluorescence recovery was monitored by
acquiring successive images during 40 s. The recovery curves were
obtained by plotting the mean fluorescence intensity as a function
of time in these two ROI, and were corrected for fluctuations in
axial position by a third ROI located into the cell, and finally
normalized to the mean value of each ROI before photobleaching. The
curves were fitted using Eq.1 which is a slightly modified 2D
diffusion model for FRAP taking into account normalization and a
mobile fraction M \citep{axe76, mat07}:
\begin{equation} \label{eqn:FrapfM}
F(t)= M \sum
\limits_{n=1}^\infty{\frac{(-K)^n}{n!}\frac{1}{1+n+2n\frac{t}{t_{1/2}}}}+(1-M)F_{0}
\end{equation}
This equation was used to its 20th order limited development for
data fitting.
Fluorescence recoveries were acquired before electric pulses and at a time t, 30 s < t < 90 s after
 electropermeabilization on the same sample using the same sequence. At this time t, more than 50\%
 of the cells were still permeabilized \citep{Rol98}.

\section{Results}
\subsection{Effects of electropermeabilization on the mobility of Rae-1 protein}
Mobility of Rae-1 protein was monitored by means of FRAP
experiments. Fitting of the recovery curve using equation
~\ref{eqn:FrapfM} lead to determination of two different parameters
which are respectively : t$_{1/2}$ which is the half-time of
recovery and M which is the mobile fraction.
\begin{figure}
  \includegraphics [scale=0.45]{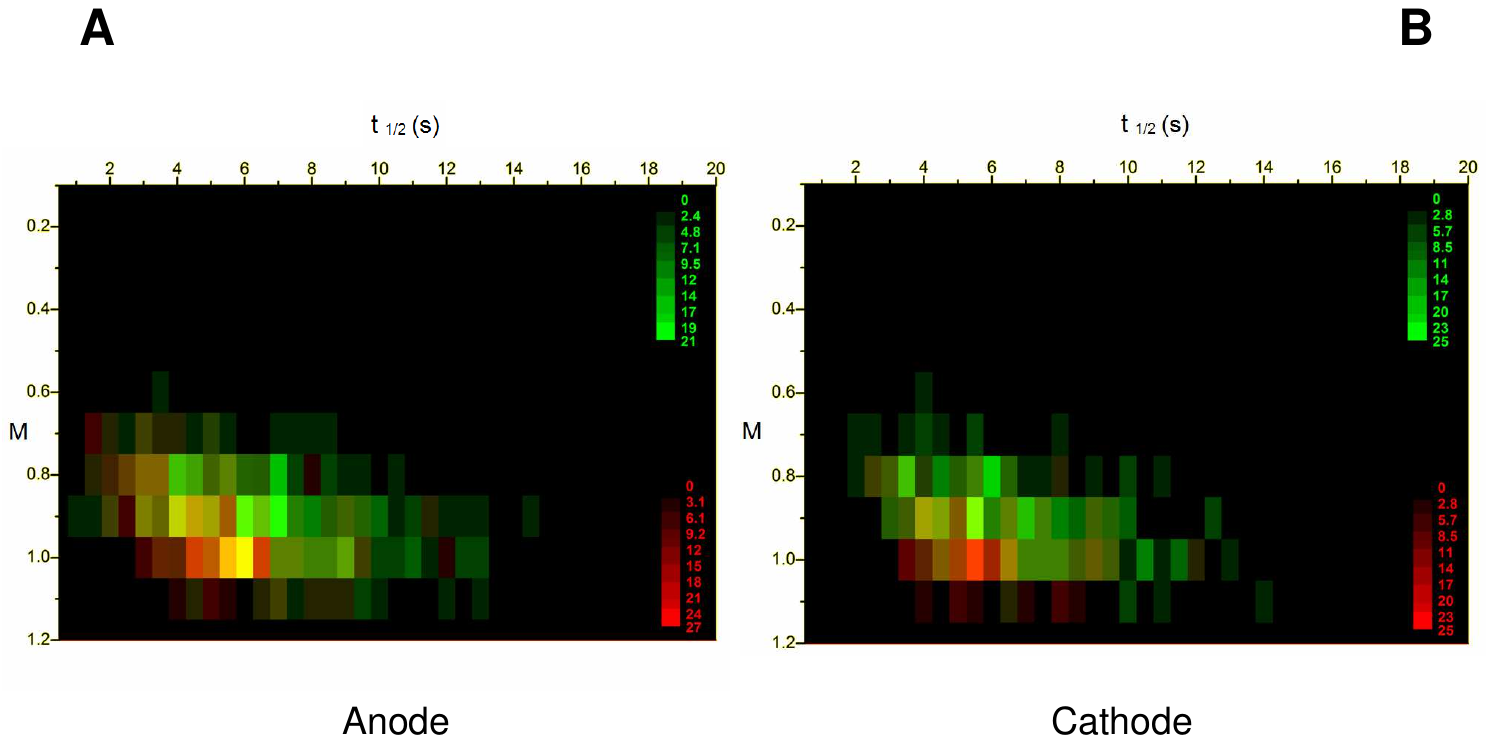}
\caption{\fontsize{8}{10}\selectfont\textbf{Distribution of
t$_{1/2}$ and M before and after electropermeabilization:}
 M was plotted as a function of t$_{1/2}$, before (in red) and after (in green) electropermeabilization
 for the cell pole facing the anode (A), or the cathode (B). Intensity
scale is dispatched on the side of each image and correspond to the
number of events in each class of (t$_{1/2}$, M) pair. }
\label{fig:1}       
\end{figure}
t$_{1/2}$ is a function
of the diffusion coefficient of Rae-1 protein, the second one is a
function of the number of mobile Rae-1 protein. In order to analyze
the mobility of Rae-1 molecule, the mobile fraction was plotted as a
function of the half-time of recovery at the anode (Figure
~\ref{fig:1}A) and the cathode (Figure ~\ref{fig:1}B) before (in
red) and after (in green) electropermeabilization, for a set of 250
different cells. The intensity of each pixel represents the number
of events belonging to each class.

\begin{table}[!htb]
\centering
\begin{tabular}{llllllll}
 &  & n&t$_{1/2}$(s)&p-value&M&p-value \\
\hline \hline
\multirow{2}{*}{Anode (1)} & before EP & \multirow{2}{*}{436} & $5.4\pm2.3$ & \multirow{2}{*}{$<10^{-4}$}& $0.88\pm0.10$ & \multirow{2}{*}{$7.10^{-4}$} \\
 & after EP&&\textbf{$7.0\pm3.2$}&&\textbf{$0.86\pm0.09$}\\
 \hline
 \multirow{2}{*}{Cathode (2)} & before EP & \multirow{2}{*}{459} & $5.6\pm2.6$ & \multirow{2}{*}{$<10^{-4}$}& $0.89\pm0.09$ & \multirow{2}{*}{$<10^{-4}$} \\
 & after EP&&\textbf{$6.9\pm3.2$}&&\textbf{$0.85\pm0.10$}\\
 \hline
 \multirow{2}{*}{Left (3)} & before EP & \multirow{2}{*}{315} & $5.3\pm1.8$ & \multirow{2}{*}{$<10^{-4}$}& $0.91\pm0.08$ & \multirow{2}{*}{$<10^{-4}$} \\
 & after EP&&\textbf{$6.2\pm2.2$}&&\textbf{$0.85\pm0.09$}\\
 \hline
 \multirow{2}{*}{Right (4)} & before EP & \multirow{2}{*}{316} & $5.8\pm1.8$ & \multirow{2}{*}{$<10^{-4}$}& $0.93\pm0.08$ & \multirow{2}{*}{$<10^{-4}$} \\
 & after EP&&\textbf{$6.7\pm2.3$}&&\textbf{$0.90\pm0.09$}\\
\hline
\end{tabular}
\includegraphics [scale=0.7]{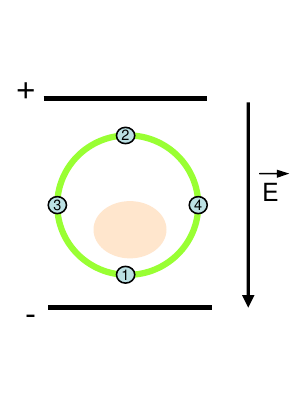}
\label{tab1} \caption{\fontsize{8}{10}\selectfont\textbf{Mean values
of t$_{1/2}$ and M before and after permeabilization and their
respective paired t-test values and localization of the measurements
in cell.} Localization of the experimental positions are depicted on
the scheme of the cell below the table.}
\end{table}

As described in Figure ~\ref{fig:1}, the mobility of Rae-1 protein
is reduced on both sides of the cell facing the electrodes (i.e.,
anode and cathode) after electropermeabilization (M slightly
decreases and t$_{1/2}$ increases). The mean values and standard
deviations of t$_{1/2}$ and M, at the pole of the cell facing the
anode and the cathode, before and after electropermeabilization, as
well as the left and right part of the equatorial level of  the cell
(perpendicular to the electric field direction) are reported in the
Table 1 using descriptive statistics. Student's t-tests has been
performed on paired values of cathode and anode showing that both
parameters scored p < 0.001 and can be considered as significantly
different before and after electropermeabilisation. Interestingly,
an increase in the of t$_{1/2}$ and a decrease of M has also been
observed in the membrane regions not facing the electrodes.

\subsection{Effects of pDNA electroinsertion on the mobility of Rae-1 protein}
\begin{figure}
\centering
  \includegraphics [scale=0.45]{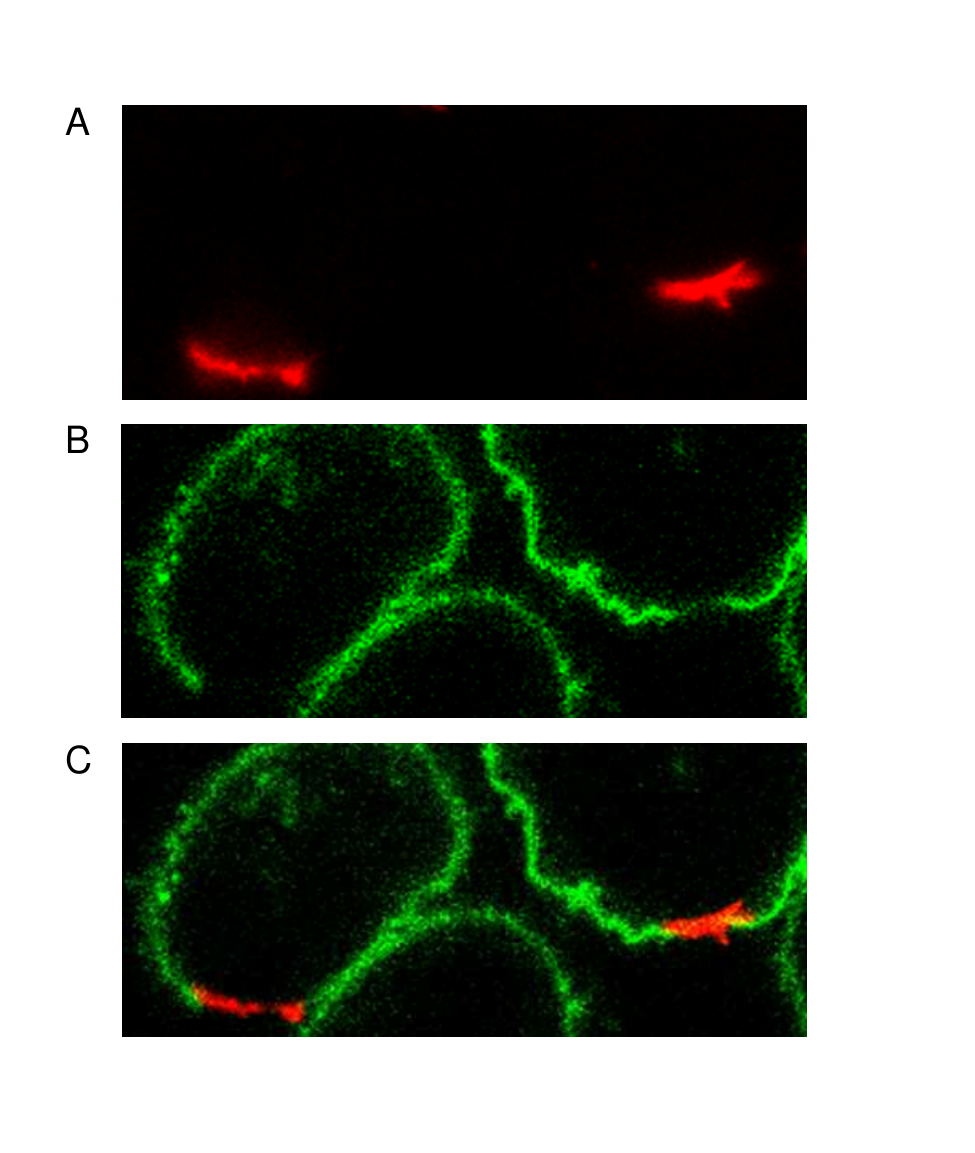}
\caption{\fontsize{8}{10}\selectfont\textbf{Images of  GFP-Rae-1 and
Cy3-pDNA after electroinsertion of Cy3-pDNA:} Cy3-pDNA is located at
the pole of the cells immediately after the end of electropulsation
(part A), covering large area at the cell surface. Fluorescence of
GFP-Rae1 is located at the plasma membrane of the CHO cells (part B)
but is clearly extincted in the area occupied by Cy3-pDNA. This is
confirmed by merging the two images (part C) where no colocalization
can be seen. }
\label{fig:2}       
\end{figure}
Photonic microscopy observations of gene electrotransfer process
reveals that pDNA molecules are found as clusters with an apparent
size close or above diffraction limit i.e. 200 nm \citep{Gol02}. We
attempted to probe the size of these pDNA clusters more accurately
by using Rae-1 dynamics. An effect on mobility of the protein may
occur since mobility is reduced by the presence of obstacles and
this reduction is proportional to the size and density of obstacles.
Unfortunately, direct image analysis of eGFP-Rae-1 expressing cells
showing an electroinsertion of Cy3-pDNA revealed that Rae-1
fluorescence was excluded from the Cy3-pDNA cluster avoiding any
possibility to perform FRAP measurements (Figure ~\ref{fig:2}).
\begin{figure}
\center
  \includegraphics [scale=0.45]{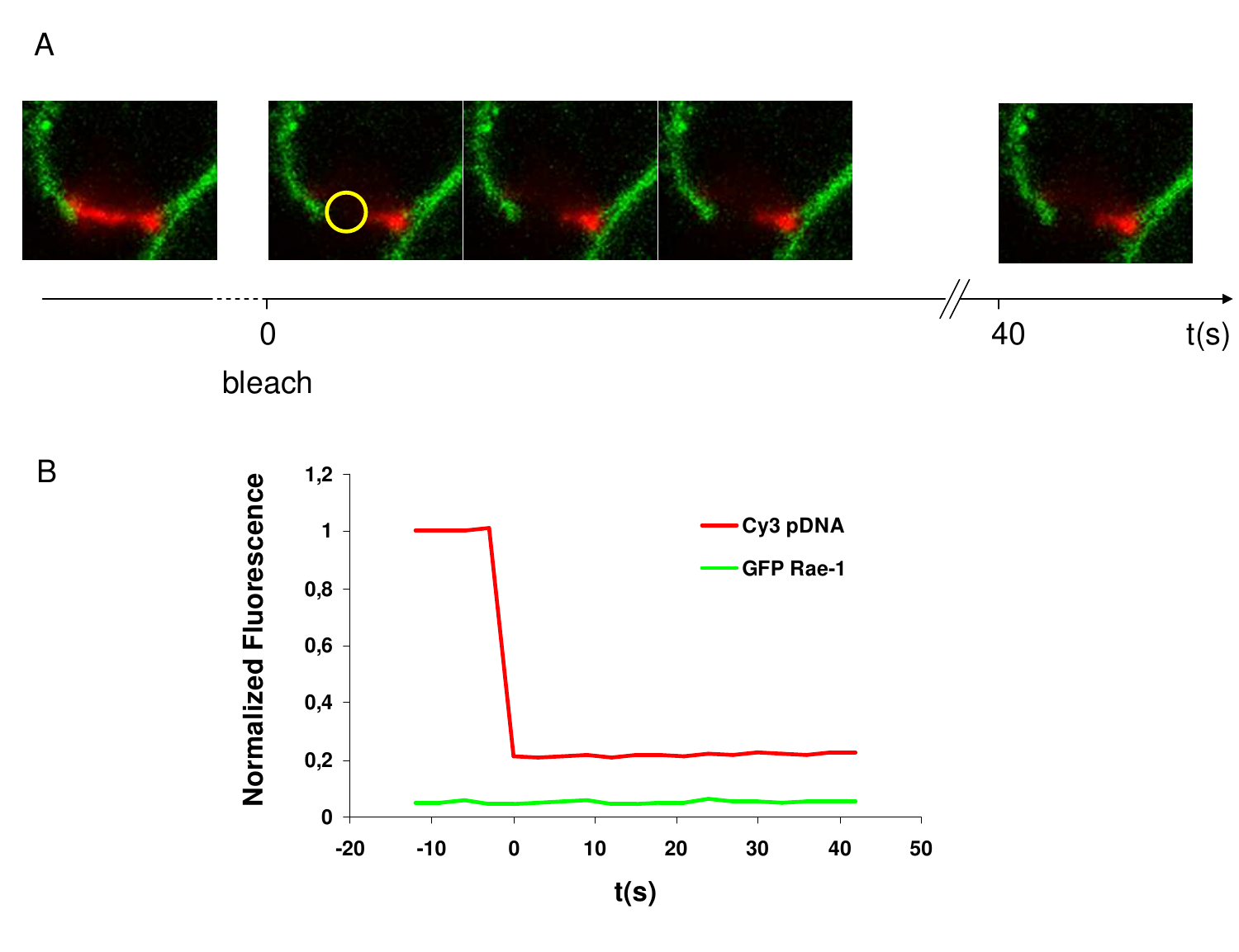}
\caption{\fontsize{8}{10}\selectfont\textbf{Photobleaching of
Cy3-pDNA after electroinsertion.} In Part A are represented images
of Cy3-pDNA (in red) and GFP-Rae-1 (in green) fluorescence acquired
during the expected recovery process. The photobleached area is
located inside the yellow circle. In Part B is shown the
fluorescence intensities integrated in the yellow circle of
Cy3-pDNA (in red) and GFP-Rae-1 (in green) normalized to the value
of Cy3-pDNA before the bleaching. These two curves clearly show that
fluorescence recovery occurred neither for Cy3-pDNA, nor for
eGFP-Rae-1, leading to the conclusion that the Cy3-pDNA is highly
immobile and that Rae-1 is totally excluded from the Cy3-pDNA
clusters located in the membrane. }
\label{fig:3}       
\end{figure}
Nevertheless, absence of measurable fluorescence does not directly
mean absence of Rae-1 proteins. Indeed, since emission spectrum of
eGFP widely overlaps absorption spectrum of Cy3, lack in
fluorescence can be due to a very efficient energy transfer of
Perrin-Forster type from eGFP to Cy3. Therefore, as shown in Figure
3, photobleaching FRET experiments were performed. The Cy3-pDNA was
photobleached using the 546 nm line of an He-Ne laser on the
confocal microscope. Occurence of FRET should result in an increase
in the fluorescence of eGFP-Rae-1 protein (Figure ~\ref{fig:3} part
B, green line). This was not the case, indicating that no FRET
occured between Cy3-pDNA and eGFP-Rae-1 protein. More interestingly,
the time evolution of the experiment immediately after the
bleaching, up to 42 s after (40 s being the total time of
acquisition in FRAP experiments of section 3.1) showed no recovery
neither for Cy3-pDNA (Figure ~\ref{fig:3} part B, red line), nor for
eGFP-Rae-1 protein, confirming a total absence of mobility of
Cy3-pDNA and indicating a total inaccessibility of eGFP-Rae-1 into
these clusters.

\section{Discussion}
\subsection{Macroscopic effects of electropermeabilization.}
\begin{figure}
  \includegraphics [scale=0.5]{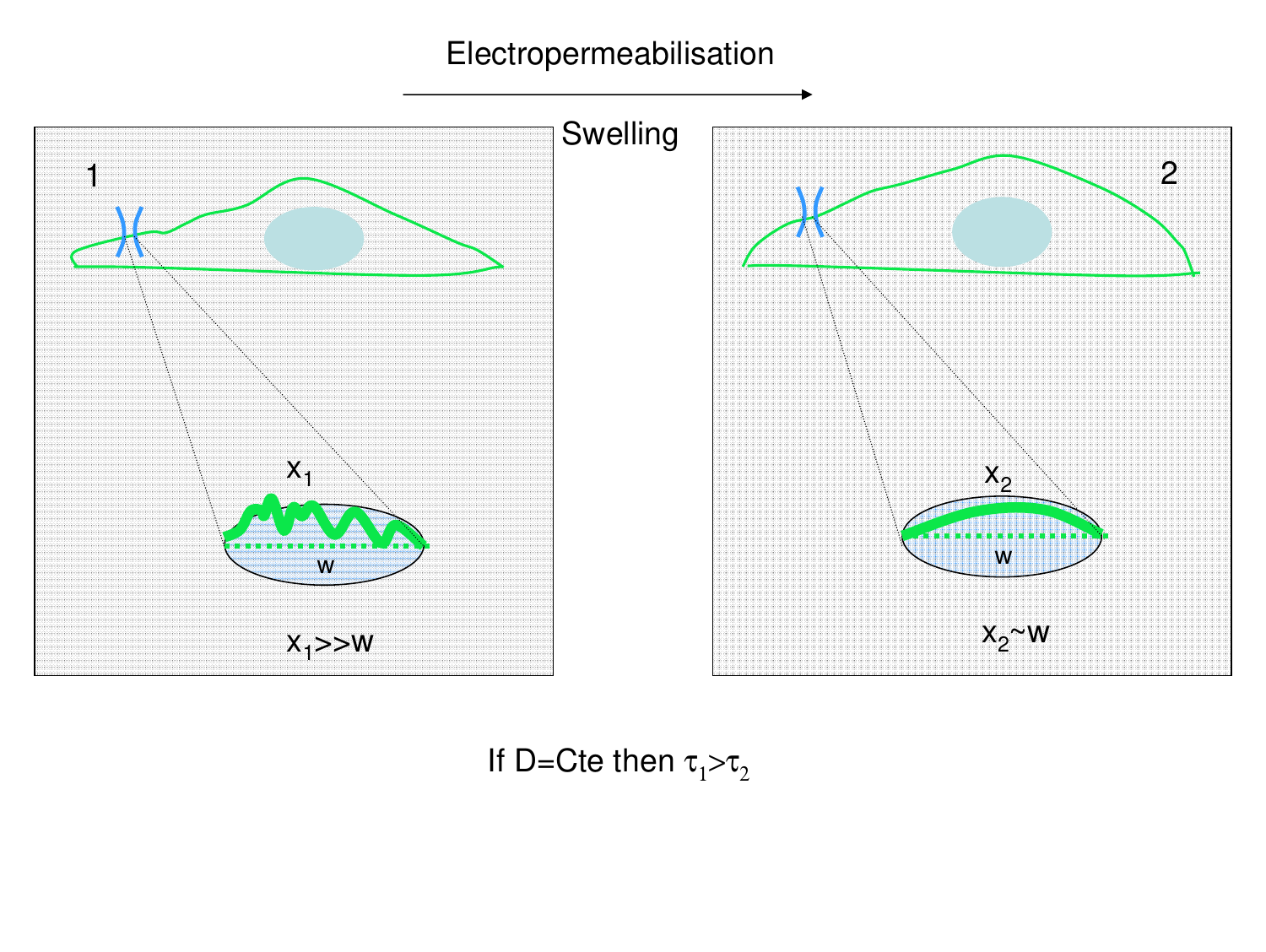}
\caption{\fontsize{8}{10}\selectfont\textbf{Incidence of the cell
swelling induced by electropermeabilization on the estimation of
recovery time in FRAP experiments:} This scheme depicts the effect
of cell swelling induced by electropermeabilization (from 1(left) to
2(right)). The swelling ends in a more tensed membrane as compared
to the resting cell. Since the laser waist is constant and assuming
that the axis of the laser is kept normal to the plane of the
membrane, it can be seen from this scheme that the area analysed
during FRAP experiments is lower after swelling than before.
Therefore, without any change in diffusion constant of the observed
molecule, the recovery time after electropermeabilization will be
lower than before.}
\label{fig:4}       
\end{figure}

Figure ~\ref{fig:4} illustrates how the osmotic swelling can locally
change the tension of plasma membrane. The FRAP recovery half-time
is the time needed for the molecule to explore the mean square of a
length defined by the waist of the laser. When applied to
measurement of lateral diffusion in cell plasma membrane, this
definition assumes that the plasma membrane is perfectly flat and
perpendicular to the laser path.
 While swelling will not affect the angle of intercept of the laser and the
 plasma membrane it can dramatically change the flatness of the plasma membrane
 (this latter becoming flatter than before swelling).
 Therefore, if a molecule diffuses with the same diffusion coefficient before and
 after electrically-mediated swelling, it will take less time to recover fluorescence
 in a tensed situation (closer to the real waist) than in a loose situation (further from the real waist).
Our results here show exactly the opposite effect, emphasizing the
existence of microscopic obstacles to free diffusion after
electropermeabilization. Moreover, our study shows that these
obstacles are also present at the equator of the cell (i.e.
locations parallel to the electrodes) one minute after
electropermeabilization. This unexpected result proves that these
obstacles are a collective signal all around the cell. This is
confirmative, but with a much faster time resolution, of the recent
work of Chopinet et al. \citep{chopinet2013a} that evidenced on
living CHO cells a rapid propagation of membrane perturbation along
the entire cell surface using AFM imaging.

\subsection{Change in the mobility of Rae-1 : relation to the permeabilized area.}
Saxton \citep{Sax82,Sax87} has theoretically showed that apparent
diffusion coefficient of a tracer decrease with the fractional area
occupied by obstacles, down to zero if obstacles are immobile at a
given area (i.e., the percolation threshold), or asymptotically if
obstacles are very mobile (as compared to the tracer), meaning that
no more percolation threshold exist. In this study, apparent
diffusion coefficient (i.e. half-time recoveries at constant given
waists) of eGFP-Rae-1 protein is decreased as much as 10\% after
electropermeabilization. This result is a confirmation of the
existence of "defects" at least in the outer leaflet of  the plasma
membrane, since the tracer, eGFP-Rae-1 protein is located in this
leaflet. This confirms previous results obtained by other approaches
such as influx of fluorescent dyes into the cytosol
\citep{Gab97,Gab99,Puc08,Rol98,Mir88,Tek94,Pra94,Pra95}. According
to Saxton's numerical simulations, a decrease of 10\% of the
apparent diffusion coefficient means that 10\% (in the case of
immobile "pores" or "defects") to 20\% (in the case of very fast
diffusing "pores" or "defects") of the total membrane area of the
cell are occupied by these "defects" or "pores". In order to compare
with experimental \citep{Gab97,Gab99,Tek01,Por11} or numerical
simulation \citep{Kra07}data available in the litterature, we have
defined an aperture angle ($0<\theta_{ap}<90^\circ$) of
permeabilisation as the half angle of the solid angle in which the
total permeabilized surface is included. In this work, if all the
obstacles were fused in a unique one, this aperture angle will be
found between 36$^\circ$ and 52$^\circ$. Portet et al. measured the
average aperture angle of pores to be 6$^\circ$ by using pDNA
translocation into giant unilamellar vesicles (GUVs) made of DOPC.
Using GUVs made of DOPC with a radius close to the CHO cells (i.e.,
15 $\mu$m) and electric fields twice our value (1.2 kV.cm$^{-1}$),
Tekle et al. \citep{Tek01} shown that up to 14\% of the total
membrane surface can be lost when pulsing, leading to a value of
44$^\circ$. Finally, Gabriel et al.\citep{Gab99} measured the
average aperture angle as the angle of the extent of the
permeabilization immediatly after the pulse by visualizing small
fluorophore entrance into the cell to be 56$^\circ$. $\theta_{ap}$
values found from these different studies exhibit discrepancies.
Many factors can account for that discrepancies amongst, which the
duration and the strength of the external applied electric field,
differences in model used (cells \citep{Gab99,Gab97} or artificial
lipid membranes \citep{Tek01,Por11}) in their radius or in the case
of differences in external and internal buffers used for
electropermeabilization. Nevertheless in \citep{Gab99} the results
have been obtained using exactly the same experimental conditions
than ours (i.e., cell type, electropermeabilization buffer and
electric field intensity range) and can therefore be directly
compared. These results exhibit a value of $\theta_{ap}$ slightly
superior to what is found here. Indeed, lifetime distribution of
"pores" or "defects" in lipid membranes spans a wide range between
microsecondes up to minutes depending on several factors (e.g.,
field intensity, conductance, lipid membrane composition, artificial
vesicules or cells) \citep{Tei05}. However, it is clear that the
number and/or size of the "pores" decreases with time after
electropermeabilization. Since our experiments have been be
conducted one minute at least after the end of pulse sequence, this
effect could account for a smaller value of $\theta_{ap}$ as
compared to the one found in \citep{Gab99,Gab97}. Alternatively,
Krassowska et al. \citep{Kra07} performed numerical simulation of
the evolution in time and space
 of pores in a spherical pure lipid vesicle exposed to an electric field.
 They simulated the normalized average area occupied by theses "pores" as a function of the external
 electric field applied to the cell. For an electric field use in this study of 0.2
 kV.cm$^{-1}$
 (that corresponds to our conditions regarding the 50 $\mu$m radius they use in their simulations) they found
 this area to be 0.07\% of the total area ( =1$^\circ$) which is far beyond our experimental results.
 Nevertheless, their values were obtained using only one pulse of 1 ms.
 If one makes this oversimplified hypothesis that the number and therefore the area occupied by the "pores"
 is a function not only of the electric field intensity, but also of the total exposure time of cell
 to an electric field. Furthermore, if one admits that the created "pores" in each impulsion do not
 reseal during our total pulse sequence, the area occupied by the pores should be 50 times more important
 than in their study, resulting in an apparent angle of ( =23$^\circ$)  which is closer to what we find here.

\subsection{Effect of pDNA electroinsertion on the lateral mobility of Rae-1.}

The apparent size of the pDNA clusters inserted into the membrane
after electropermeabilization (larger than 200 nm in diameter,
\citep{Gol02,Phe05}) should lead to a drastic increase in the
half-time recovery (t$_{1/2}$) and a drastic decrease in the mobile
fraction (M) due to eGFP-Rae-1 proteins trapped in this pDNA
clusters. Unfortunately the very low level or the total absence of
fluorescence of Rae-1 in the pDNA clusters at the membrane did not
permit to perform efficient FRAP measurements.

This absence of fluorescence could be of two different origins :

\begin{itemize}
\item Loss in fluorescence of the eGFP-Rae-1 protein due to very
efficient energy transfer of Perrin-Förster type (FRET) between eGFP
fused Rae-1 protein and Cy-3 labeling the pDNA.
\item Exclusion of Rae-1 molecules from the pDNA clusters.

\end{itemize}

Photobleaching FRET experiments have been performed in order to
discriminate between this two possibilities. The absence of recovery
in the fluorescence of Rae-1 proteins when the Cy-3 molecules are
bleached clearly shows that eGFP-Rae-1 proteins are totally excluded
from pDNA clusters induced by electropermeabilization. On the other
hand, no recovery occurs for pDNA fluorescence, confirming that the
pDNA clusters are highly immobile, as previously described
\citep{Esc11}. This results is in favor of a pDNA aggregate that
exclude some of the membrane components, if not all, to its
periphery.

\section{Conclusion}

In order to sense at the molecular scale the effect of
electropermeabilisation and pDNA electroinsertion on the plasma
membrane lateral organisation, mobility of a tracer (Rae-1 protein)
has been assessed in living cells. This study show that membrane
reporters can sense the defaults induced by electropermeabilisation
and estimate correctly the surface permeabilized during field
exposure. More interestingly, this study shows that these defaults
can propagate rapidly allover the surface of the plasma membrane.
The local creation of a permeabilized cap on the cell surface
triggers a global cellular effect. Finally, attempting to perform
such an approach to characterize the nature of pDNA clusters
immediately after electroinsertion, we have shown that these
clusters were highly dense since they did not allow Rae-1 protein to
penetrate into them.

%
%

\section {Acknowledgements}
\fontsize{10}{12}\selectfont\textit{This work has been performed in
collaboration with the "Toulouse Réseau Imagerie" core IPBS facility
(Genotoul, Toulouse, France), which is supported by the Association
Recherche Cancer, Region Midi Pyrenees, the European union (FEDER)
and Grand Toulouse cluster. This research project was conducted in
the scope of EBAM European Associated Laboratory and of the COST
Action TD1104. JM Escoffre was the recipient of an allocation de
recherche du Ministère de l'Enseignement Supérieur et de la
Recherche. The authors are grateful to Dr. M. Golzio for critical
reading of the manuscript. C. Favard is a membership of CNRS
consortii GDR2588 "MIV" and GDR3070 "CellTiss".}

\bibliographystyle{spbasic}      
\bibliography{JMFRAP2}   

%
%

\end{document}